\documentclass[letter]{aa}  

\usepackage{graphicx}
\usepackage{txfonts}
\usepackage[colorlinks=true,allcolors=blue]{hyperref}
\usepackage{xcolor}
\usepackage{textgreek}
\usepackage{orcidlink}
\usepackage{bbding}
\usepackage{multirow}
\begin{document} 

\newcommand{\solarM}{\,\mathrm{M}_\odot}
\newcommand{\solarL}{\,\mathrm{L}_\odot}
\newcommand{\E}[1]{\times10^{#1}}
\newcommand{\nH}{\,\mathrm{cm}^{-2}}
\newcommand{\ergcms}{\,\mathrm{erg}\,\mathrm{cm}^{-2}\,\mathrm{s}^{-1}}
\newcommand{\asec}{\,\mathrm{arcsec}}
\newcommand{\amin}{\,\mathrm{arcmin}}
\newcommand{\magnitude}{\,\mathrm{mag}}
\newcommand{\ergs}{\,\mathrm{erg}\,\mathrm{s}^{-1}}
\newcommand{\pdotunit}{\,\mathrm{s}\,\mathrm{s}^{-1}}
\newcommand{\LX}{L_\mathrm{X}}
\newcommand{\LEdd}{L_\mathrm{Edd}}
\newcommand{\pspin}{P_\mathrm{spin}}
\newcommand{\porb}{P_\mathrm{orb}}
\newcommand{\asini}{a_\mathrm{x}\sin i}
\newcommand{\FX}{F_\mathrm{X}}
\newcommand{\apm}[2]{_{-#1}^{+#2}}
\newcommand{\xmm}{\emph{XMM-Newton}}
\newcommand{\swift}{\emph{Swift}}
\newcommand{\chandra}{\emph{Chandra}}
\newcommand{\erosita}{\emph{eROSITA}}
\newcommand{\nustar}{\emph{NuSTAR}}
\newcommand{\askap}{ASKAP}
\newcommand{\ep}{\emph{Einstein Probe}}
\newcommand{\fxt}{EP/FXT}
\newcommand{\src}{ASKAP\,J1745}
\newcommand{\srclong}{ASKAP\,J174508.9-505149}
\newcommand{\salt}{\emph{SALT}}
\newcommand{\rxte}{\emph{RXTE}}
\newcommand{\citerose}{Rose et al. (2026, in press.)}
\newcommand{\citeprose}{(Rose et al., 2026 in press.)}
\newcommand{\citealprose}{Rose et al., 2026 in press.}
\newcommand{\matt}[1]{\textcolor{red}{#1 - MI}}
\newcommand{\wyl}[1]{\textcolor{purple}{#1 - YW}}
\newcommand{\rev}[1]{\textbf{#1}}

   \title{The X-ray emission of the long-period transient and accreting cataclysmic variable ASKAP J174508.9-505149}
   \titlerunning{The X-ray emission of the LPT/CV ASKAP J174508.9-505149}
   \authorrunning{Imbrogno et al.}

   \author{M. Imbrogno
          \inst{1,2,3}
          \orcidlink{0000-0001-8688-9784}
          \and
          M. Veresvarska
          \inst{1,2}
          \orcidlink{0000-0002-0146-3096}
          \and
          Y. L. Wang
          \inst{4,1,2,5}
          \orcidlink{0009-0009-1721-3663}
          \and
          N. Rea
          \inst{1,2}
          \orcidlink{0000-0003-2177-6388}
          \and 
          F. Coti Zelati
          \inst{1,2}
          \orcidlink{0000-0001-7611-1581}
          \and
          K. Rose
          \inst{6,7}
          \orcidlink{0000-0002-7329-3209}
          \and
          J. Pritchard
          \inst{7}
          \orcidlink{0000-0003-1575-5249}
          \and
          D. de Martino
          \inst{9}
          \orcidlink{0000-0002-5069-4202}
          \and
          S. Scaringi
          \inst{9,10}
          \orcidlink{0000-0001-5387-7189}
          \and
          Z. Wang
          \inst{11}
          \orcidlink{0000-0002-2066-9823}
          \and
          D.~L. Kaplan
          \orcidlink{0000-0001-6295-2881}
          \inst{12}
          }

   \institute{Institute of Space Sciences (ICE, CSIC), Campus UAB, Carrer de Can Magrans s/n, E-08193 Barcelona, Spain
        \and 
            Institut d'Estudis Espacials de Catalunya (IEEC), 08860 Castelldefels (Barcelona), Spain 
        \and 
            INAF -- Osservatorio Astronomico di Roma, via Frascati 33, I-00078 Monte Porzio Catone (RM), Italy
        \and 
            National Astronomical Observatories, Chinese Academy of Sciences, 20A Datun Road, Beijing 100101, China
        \and 
            School of Astronomy and Space Science, University of Chinese Academy of Sciences, 19A Yuquan Road, Beijing 100049, China
        \and 
            Sydney Institute for Astronomy, School of Physics, The University of Sydney, Sydney, 2006, NSW, Australia
        \and 
            Australia Telescope National Facility, CSIRO, Space \& Astronomy, PO Box 76, Epping, 1710, NSW, Australia
        \and 
            ARC Centre of Excellence for Gravitational Wave Discovery (OzGrav), Hawthorn, VIC, Australia
        \and 
            INAF-Osservatorio Astronomico di Capodimonte, INAF, Salita Moiariello 16, Naples, 80131, Italy
        \and 
            Centre for Extragalactic Astronomy, Department of Physics, Durham University, South Road, Durham, DH1 3LE, UK
        \and 
            International Centre for Radio Astronomy Research, Curtin University, Bentley, WA 6102, Australia
        \and 
            Department of Physics, University of Wisconsin-Milwaukee, P.O. Box 413, Milwaukee, WI 53201, USA
             }

   \date{Received MONTH XX, YYYY; accepted MONTH XX, YYYY}

  \abstract{
  Long-period transients (LPTs) challenge our knowledge of the mechanism producing radio periodic pulsations in compact objects. Some LPTs have been associated with systems hosting a white dwarf and a low-mass star in a detached binary. Recently, a new LPT (\srclong) has been classified as an accreting cataclysmic variable (CV). In the present letter, we report on the detailed study of the X-ray variability of \srclong\ as observed by \xmm\ and \ep\ between September 2025 and May 2026. Simultaneous optical and radio observations are also presented. We studied the timing variability of the source, and estimated an X-ray periodicity of $P=4868(22)$\,s, consistent with radio and optical periods. We also observe the same periodicity in the hardness ratio extracted from the \xmm\ observation, peaking at the minimum of the modulation.
  A long-term modulation is also present in the X-rays and in the B-band photometry, but it is poorly constrained by the current data set. Spectral X-ray analysis shows the presence of a black-body component ($\sim$0.1\,keV), a collisionally ionized plasma ($\sim$15\,keV), and an absorption feature at 0.77 keV (possibly due to Oxygen-VII). This is the third LPT detected in the X-ray band, the second with a detected X-ray periodicity and variable X-ray emission, and the first conclusively recognised as an accreting magnetic CV.}

   \keywords{
       White dwarfs -- cataclysmic variables -- X-rays: individuals: ASKAP\,J174508.9-505149 -- accretion, accretion disks     
        }

   \maketitle
%

\section{Introduction}\label{sec:introduction}

Long-period transients (LPTs) are a class of recently identified radio sources characterised by periodic radio bursts, with periodicities spanning from minutes to hours \citep[for a review, see][]{Rea2026}. They show bright (up to a few tens of Jy) radio pulses with strong linear and circular polarizations \citep[up to 100\%; see e.g.][]{Caleb2024,Bloot2025}, hinting at a coherent radio emission process, possibly involving magnetic fields. \cite{Rea2026} identified 12 LPTs (see their Table~1), but the number is rapidly growing\footnote{For an updated list, see \url{https://vast-survey.org/LPTs/}}. 

The nature and the emission mechanism of LPTs are puzzling. The high polarization and the coherent emission initially pointed towards isolated, strongly magnetised neutron stars slowed down by fall-back accretion at birth \citep[see e.g.][]{Hurley-Walker2022,Ronchi2022}. However, the long periodicity would put LPTs in the so-called "death valley" of the $P-\dot{P}$ diagram, where no coherent emission is expected by a typical radio-pulsar pair-production process \citep{Hurley-Walker2023, Suvorov2023,Rea2024}. The observed spin periods and expected population are more consistent with white dwarf (WD) emitters, but the higher luminosities and polarisation in LPTs are at odds with the known AR Sco-like objects \citep{Marsh2016,Buckley2017,Pelisoli2024,CastroSegura2025}. More recently, two LPTs have been recognized as WD-M dwarf binaries \citep{Hurley-Walker2024,deRuiter2025,Rodriguez2025a,Rodriguez2026}, thus models involving binary systems hosting a magnetised WD and a low-mass companion have gained significant traction \citep[see e.g.][and references therein]{Horvath2026}. The detection of X-ray counterparts to LPTs ASKAP\,J1832-0911 \citep{Wang2025} and ASKAP\,J144834-685644 \citep{Anumarlapudi2025}, and the variability and periodic X-ray modulation of the former, has opened the interesting possibility of at least some LPTs being accreting systems \citep[CVs; for reviews, see][]{Warner1995,Page2022,Scaringi2026}, but no LPT-like radio emission was conclusively detected from an accreting CV until now. 

Accreting CVs with strongly magnetic WDs are typically divided in two main classes. When the WD magnetic field $B\gtrsim10^{7-8}$\,G, the WD spin and orbit of the order of hours are synchronised, and material is accreted along magnetic field lines as no accretion disc is formed. These systems are usually referred to as "polars". On the other hand, when $B\simeq10^5-10^7$\,G the spin ($\sim$\,minutes) and orbit ($\sim$\,hours) are not (yet) synchronised. These systems are known as intermediate polars (IPs). In such cases, an accretion disc/ring generally forms, but is truncated at the WD magnetospheric radius \citep{Ferrario2015}. 

\srclong\ (\src) was discovered during an untargeted search for circularly polarised sources with the Australian SKA Pathfinder radio telescope \citep[ASKAP;][]{Hotan2021}. It is the first LPT whose periodicity $P= 4841.89\apm{0.14}{0.12}$\,s has been detected in the radio (ASKAP, MeerKAT), optical (SOAR), and X-ray band (\ep, EP), and the first LPT conclusively identified as an accreting WD-M dwarf binary \citep{Rose2026}. Here we report the X-ray analysis of \src\, as observed by EP and \xmm, between September 2025 and May 2026. Furthermore, we present simultaneous radio (ASKAP) and B-band (\xmm/OM) observations.

\section{Observations and data analysis}\label{sec:ObsDataReduction}

We requested a series of target of opportunity observations encompassing X-ray, optical, and radio bands (Table~\ref{tab:XrayObs}) to monitor the multi-wavelength flux, timing and spectral evolution of \src. The reported X-ray net exposure times take into account flare filtering. The details of the data reduction are reported in Appendix~\ref{appendix:datareduction}. Unless otherwise stated, the errors reported in this letter correspond to $1\sigma$ (68.3\%) confidence levels. To compute the unabsorbed luminosities, we considered a distance $d\simeq0.57$\,kpc, estimated through \textit{Gaia} parallax and supported by the M-dwarf companion magnitude \citep{Rose2026}. We note, however, that there is still some uncertainty on \src\ distance, which according to photogeometric estimates can be up to $d\simeq6.5$\,kpc \citep{Rose2026}.  

\subsection{X-ray data analysis}

\begin{figure*}
    \centering
    \includegraphics[width=0.67\columnwidth]{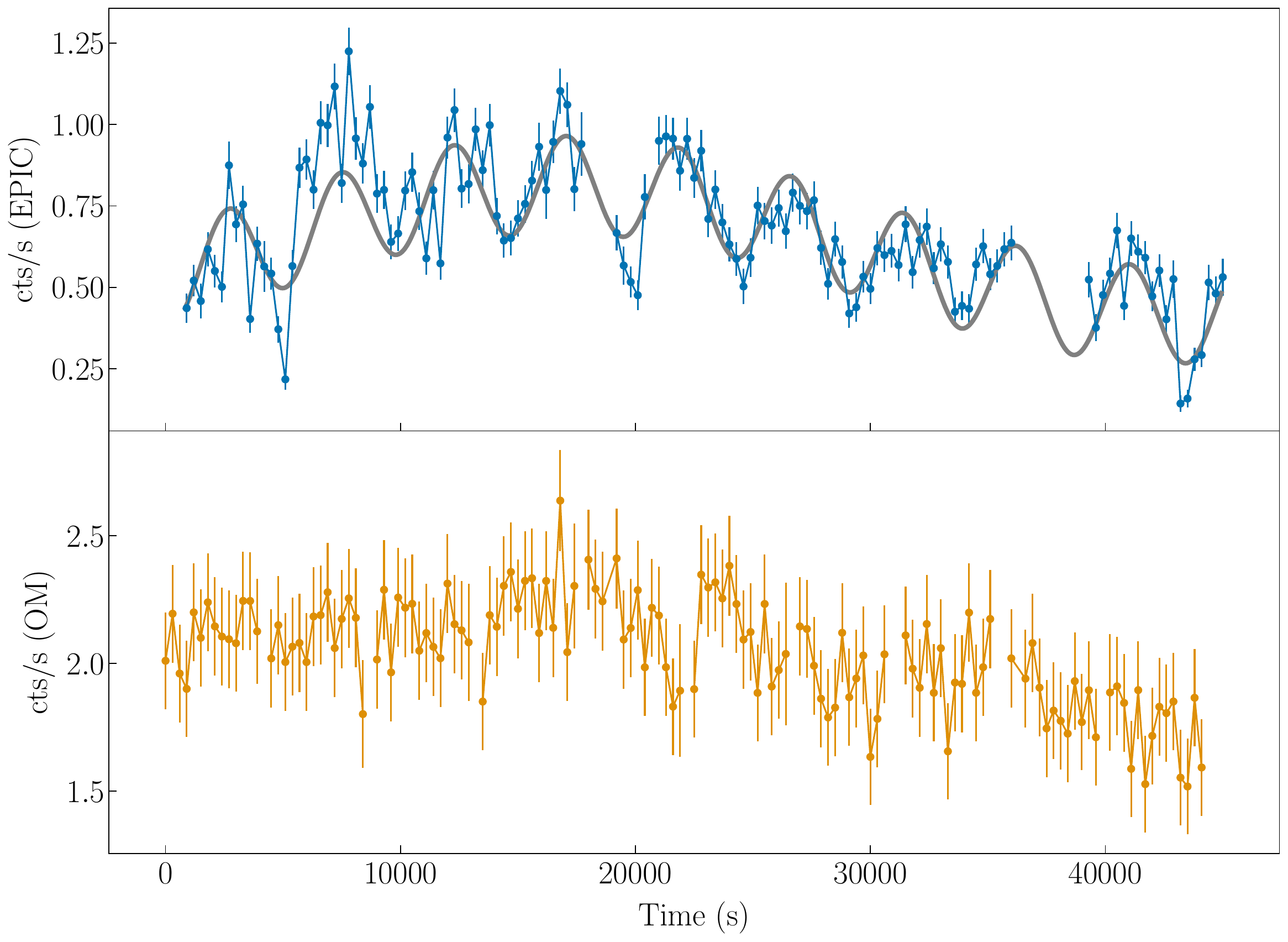}
    \includegraphics[width=0.67\columnwidth]{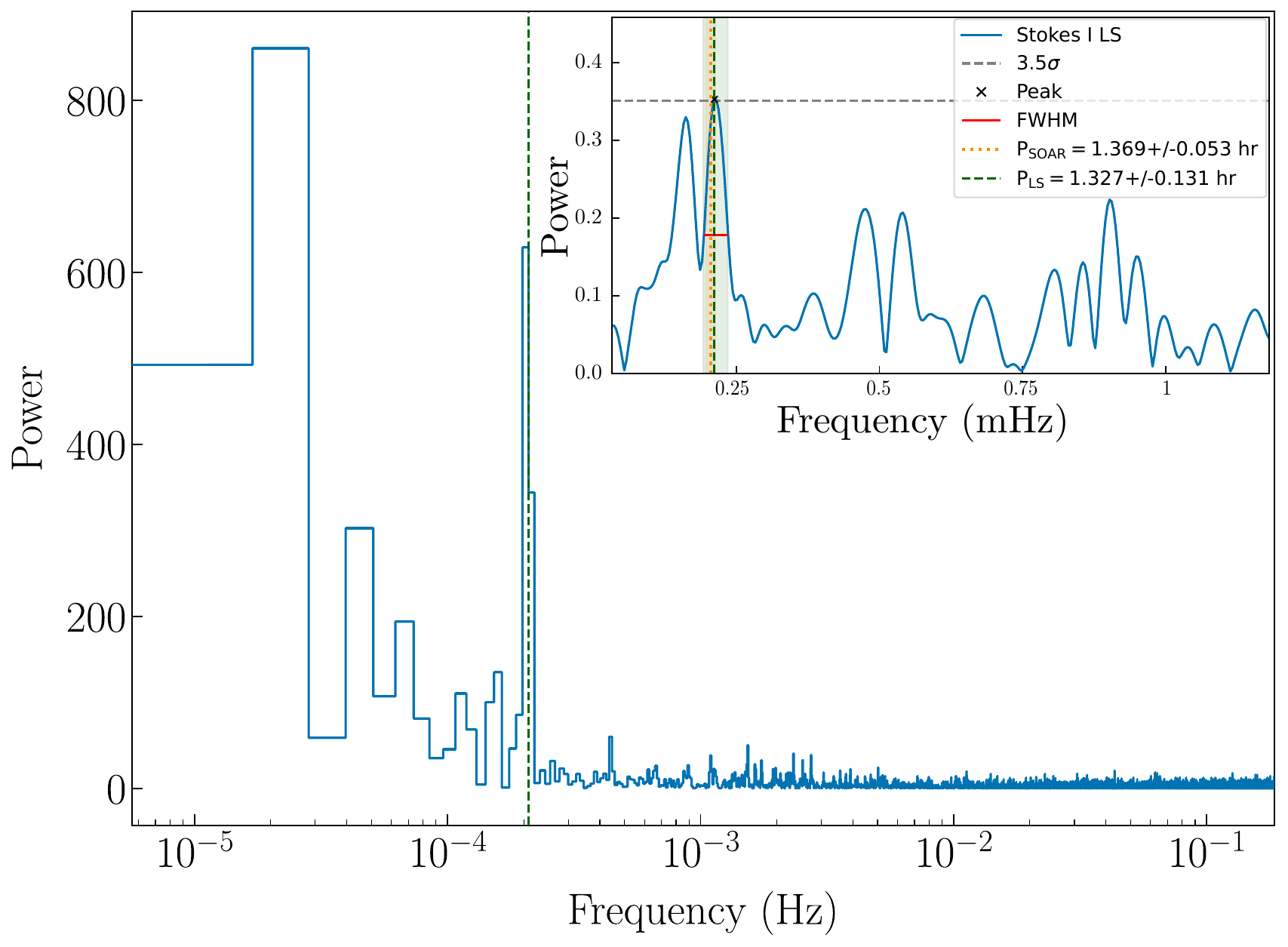}
    \includegraphics[width=0.67\columnwidth]{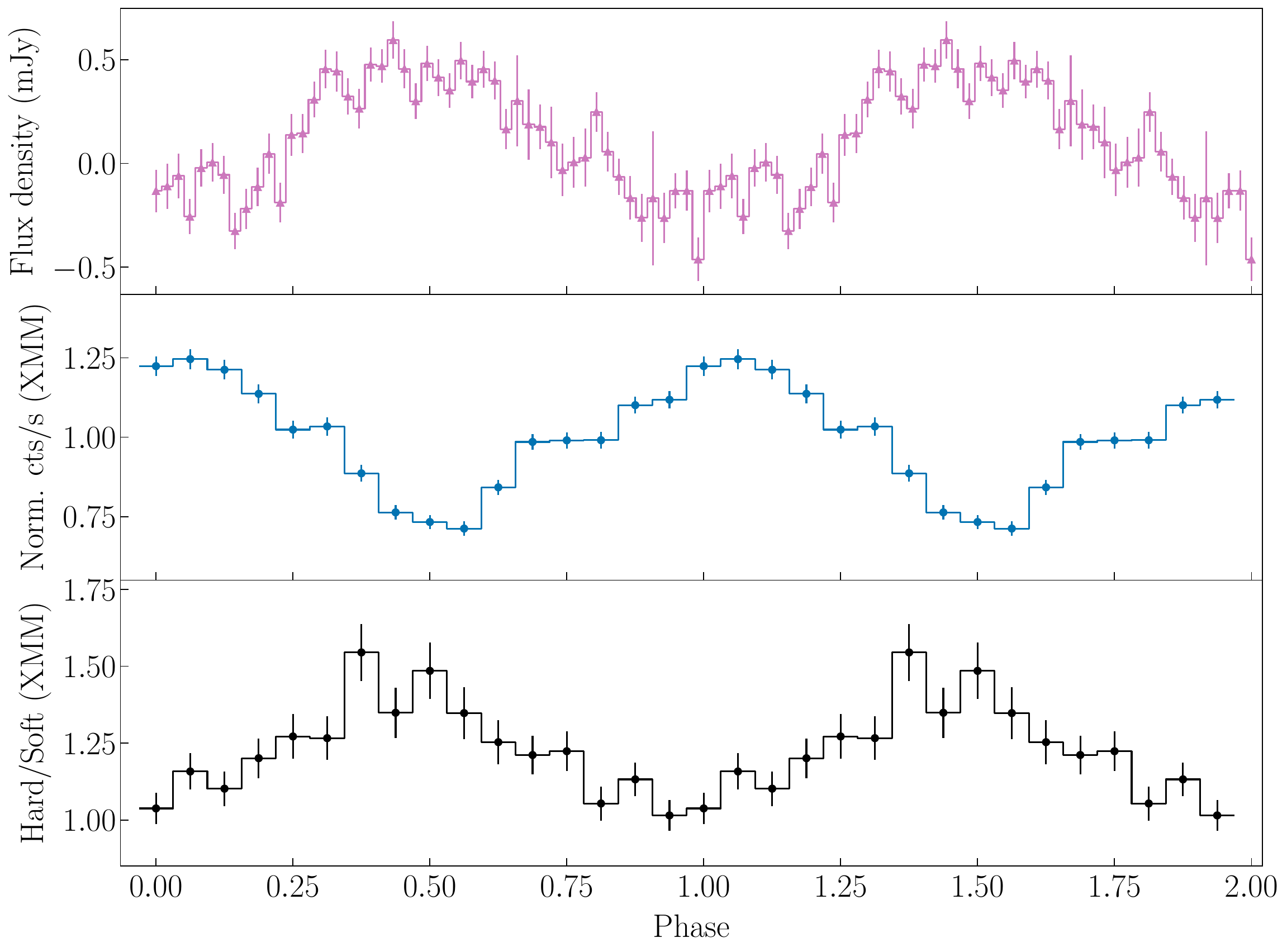}
    \caption{\textit{Left}: 0.3--10\,keV \xmm\ light curve (top) and (bottom) optical B-band (effective wavelength 450\,nm; width 105\,nm) OM light curve. In grey, the best-fit model. The bin time is 300\,s. 
    \textit{Middle}:0.3--10\,keV PDS of the \xmm\ observation (blue line), computed combining data from EPIC-pn/MOS cameras, and using the Leahy normalization. The dark green, dashed line shows the frequency of the radio signal detected by \cite{Rose2026}. In the inset: radio LS periodogram of the ASKAP observation 77513 with the detection of the signal (amplitude $\gtrsim3.5$ times the standard deviation of the radio LS power). \textit{Right}: Phase-folded profile of \src\ X-ray and radio light curves ($P=4868$\,s, reference epoch MJD 60952). Top panel: ASKAP observation. Middle panel: \xmm\ observation. Bottom panel: folded hardness ratio of the \xmm\ observation, computed as the ratio of the hard (1.5--10\,keV) count rate over the soft (0.3--1.5\,keV) count rate.
    }
    \label{fig:lcfit}
\end{figure*}

In Fig.~\ref{fig:pds}, we show the FXT1, FXT2, and FXT3 light curves, while in the left panel of Fig.~\ref{fig:lcfit} we show the \xmm\ light curve in the 0.3--10\,keV band. The modulation is clearly visible in all light curves, especially in the long \xmm\ exposure. In the middle panel of Fig.~\ref{fig:lcfit}, we show the power density spectrum (PDS) in the 0.3--10\,keV band of the \xmm\ data, in which a peak is visible at a frequency of $\simeq2\E{-4}$\,Hz. 

We report the details of the timing analysis in Appendix~\ref{appendix:boot}. Through phase-fitting, we estimated a period $P=4868(22)$\,s. In the right panel of Fig.~\ref{fig:lcfit}, we show the folded \xmm\ light curve in the 0.3--10\,keV band and the \askap\ folded light curve. The EP folded light curve is shown in Fig.~\ref{fig:epprofile}. All the light curves are folded with the same timing solution ($P=4868$\,s, reference epoch MJD 60952). The radio and X-ray profiles show a 180$^\circ$ shift in phase. For the \xmm\ observation, we estimate a pulsed fraction (defined as the semi-amplitude of the sinusoid divided by the source average count rate) of $PF_\mathrm{XMM}=23(1)\%$. For the combined \fxt\ observations, we estimate a pulsed fraction $PF_\mathrm{EP}=22(3)\%$. In Fig.~\ref{fig:lcfit}, we also show the folded hardness ratio of the \xmm\ data, defined as the ratio of the hard (1.5--10\,keV) counts over the soft (0.3--1.5\,keV) counts. The hardness ratio shows hardening at the minimum of the flux modulation. By folding the soft and hard \xmm\ light curves, we derived a pulsed fraction of $PF_\mathrm{soft}=28(1)\%$ and $PF_\mathrm{hard}=17(1)\%$, respectively.

The \xmm\ light curve also shows flux modulation on a longer timescale. To describe this additional component, tentatively reported by \cite{Rose2026} at timescales $P_\mathrm{long,radio}\simeq8$\,hr$\simeq29$\,ks, we modelled the light curve with two sinusoidal functions (left panel of Figure~\ref{fig:lcfit}).  We considered a constant component (to account for the mean count rate) and two sinusoidal components, one with period fixed at the phase-fitting period $P=4868$\,s, and one to model the long-term modulation observed in the light curve. For the latter, which significantly improves the fit (significance $>$5$\sigma$), we derive a period $P_\mathrm{long}=52(2)$\,ks, approximately twice $P_\mathrm{long,radio}$. However, we note that $P_\mathrm{long}$ is very close to the net exposure time ($\simeq44.7$\,ks).

To analyse the X-ray spectra of \src, we used \textsc{XSPEC} v12.15.0 \citep{Arnaud1996}. We adopted the element abundances and cross sections of \cite{Wilms2000} and \cite{Verner1996}, respectively. The uncertainties reported for the spectral parameters correspond to 90\% confidence ranges. We adopted the pseudo-model \texttt{cflux} to compute the unabsorbed fluxes. Using the 3D NH-Tool\footnote{\url{http://astro.uni-tuebingen.de/nh3d/nhtool}} \citep{NHTool2024}, we found that along the line of sight of \src\ the Galactic absorption grows from $1\E{21}\nH$ to $2\E{21}\nH$ between 0.5 and 6.5\,kpc. We therefore assumed the average value $1.5\E{21}\nH$. Given the lower statistics of the \fxt\ data, we modelled the spectra from observations FXT1, FXT2, FXT3, FXT4, and FXT5 with a simple absorbed power-law model (\texttt{tbabs$\times$powerlaw} in \textsc{XSPEC} syntax), with the absorption fixed at $N_\mathrm{H}=1.5\E{21}\nH$. The fit is satisfactory in all five observations ($\chi^2/\mathrm{dof}\simeq1-1.2$). 

\begin{figure}
    \centering
    \includegraphics[width=0.85\columnwidth]{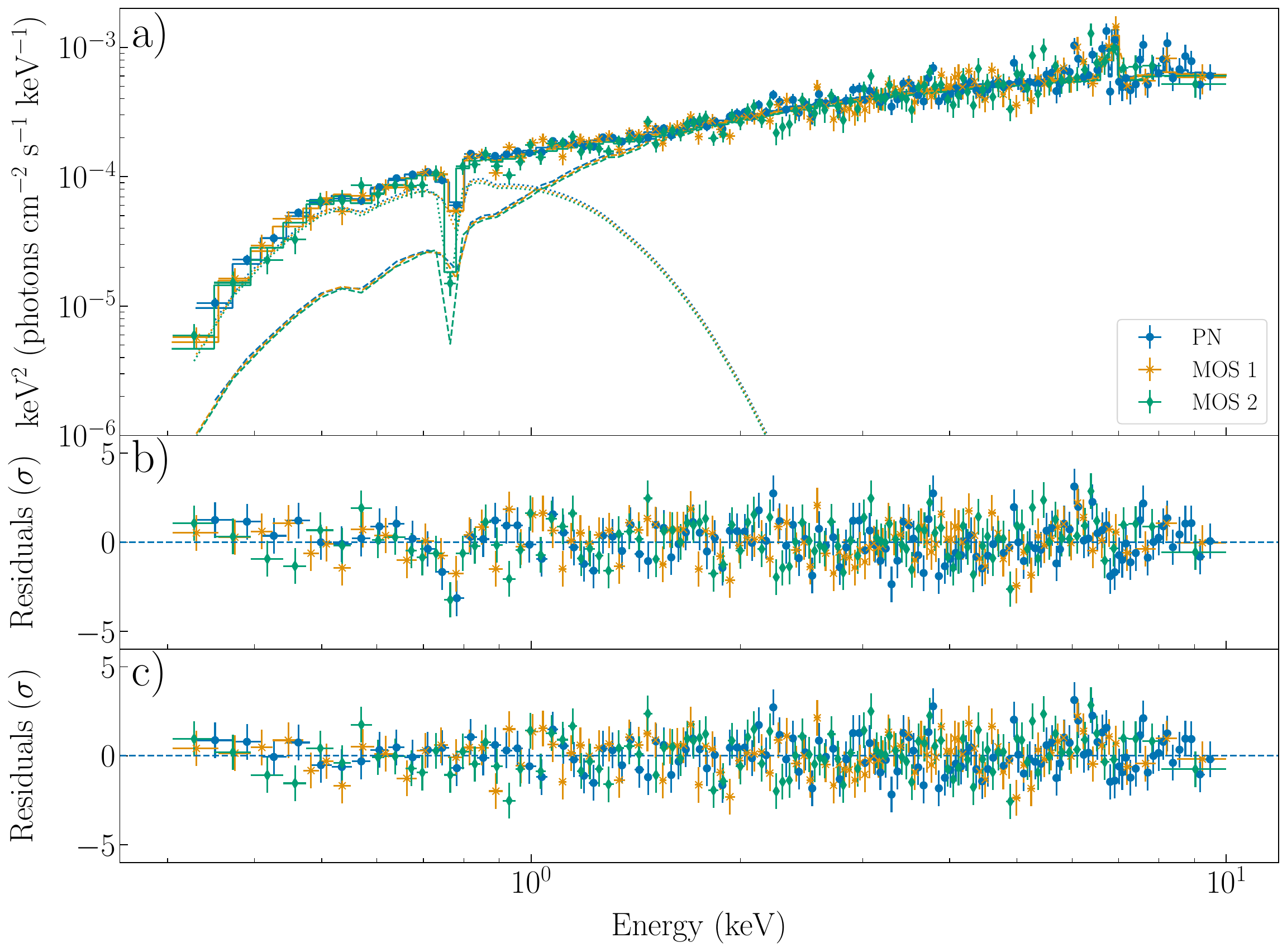}
    \caption{\xmm\ energy spectrum of \src. Panel a: Unfolded EPIC-pn/MOS spectra and best-fit model. The dashed (dotted) line shows the APEC (black-body) component of the model. Panel b: Residuals, in units of standard deviation, without the \texttt{gabs} component. Panel c: Residuals, in units of standard deviation.
    }
    \label{fig:xmmspectrum}
\end{figure}

 To model the \xmm\ energy spectrum, we considered a \texttt{constant} component to take into account inter-calibration differences between the cameras, freezing the PN constant to 1. We included a \texttt{tbabs} component to take into account the Galactic absorption. The hardness evolution shown in Fig.~\ref{fig:lcfit} indicates an intrinsic absorption effect of local matter whose partial covering changes during the period. Therefore, we also included a \texttt{tbpcf} component. The spectrum is well-described by an \texttt{apec} plus \texttt{bbodyrad} model. There are structured residuals at energies $\simeq0.8$\,keV (panel b of Fig.~\ref{fig:xmmspectrum}), thus we added a \texttt{gabs} component. 

\begin{figure}
    \centering
    \includegraphics[width=0.85\columnwidth]{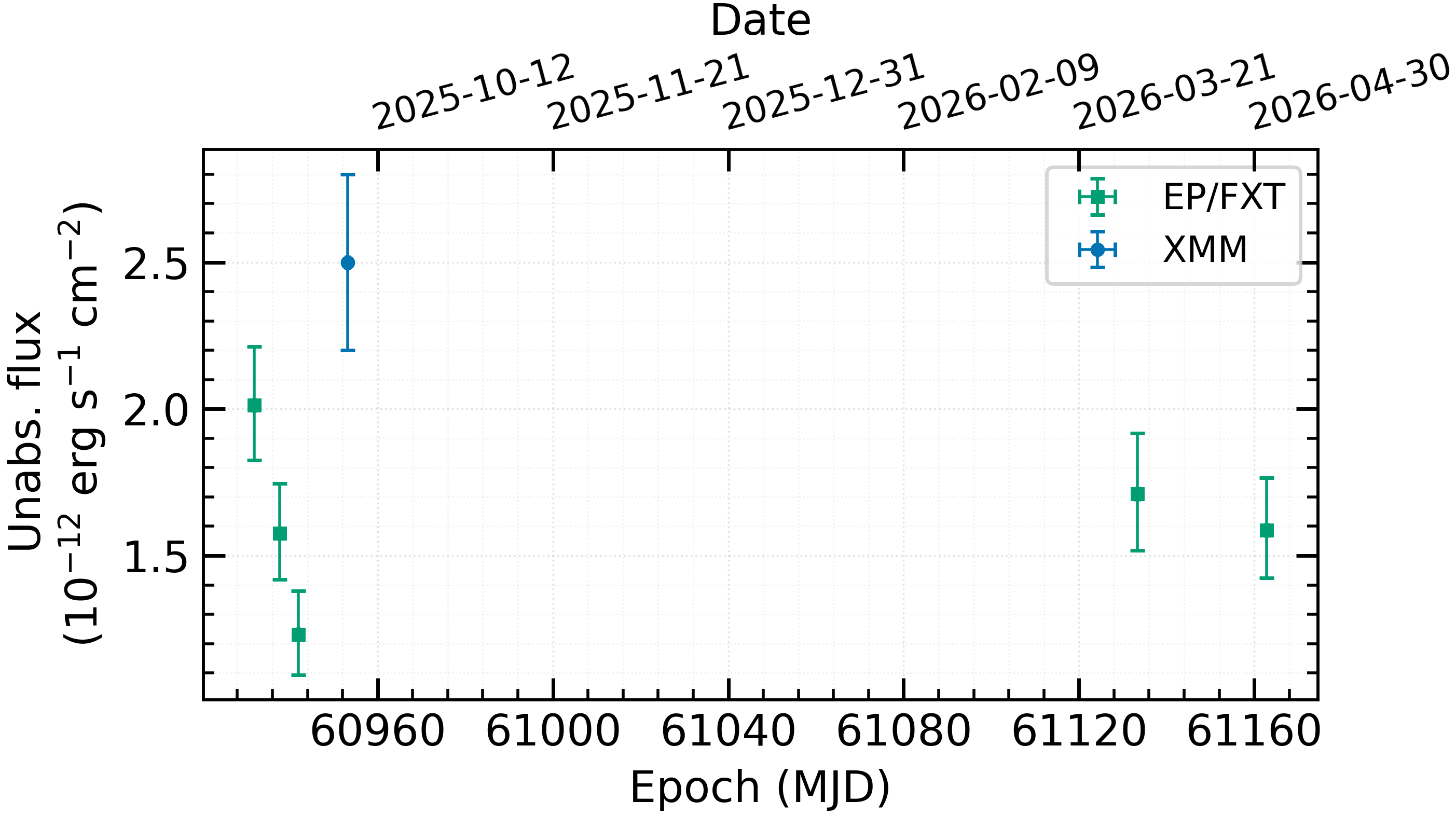}
    \caption{Evolution of the 0.5--10\,keV unabsorbed flux. Error bars denote 90\% confidence level uncertainties.}
    \label{fig:FXTevolution}
\end{figure}

 In Fig.~\ref{fig:xmmspectrum}, we show the unfolded spectrum from the \xmm\ observation, together with the best-fit model and residuals. We report the values of the parameters of our best-fit model in Table~\ref{tab:spec}. For the absorption line at low energies, we estimate a centroid energy of $0.77(1)$\,keV, consistent with absorption from O\,VII. We also find a plasma temperature for the \texttt{apec} component $kT_\mathrm{apec}=15\apm{3}{4}$\,keV and a blackbody temperature for the \texttt{bbodyrad} component $kT_\mathrm{bb}=0.14\apm{0.01}{0.03}$\,keV.  We estimated a 0.5--10\,keV unabsorbed flux $F_\mathrm{0.5-10\,keV}=2.5(3)\E{-12}\ergcms$, showing a re-brightening relative to FXT3, and indicating fluctuations in the flux evolution (Figure~\ref{fig:FXTevolution}).

\subsection{Optical data analysis}

We detected the optical counterpart of \src\ with the OM. The position of the source is consistent with that of the source Gaia DR2 5946454415417964032 (Gaia4032), as \cite{Rose2026} concluded. For the OM data, we derived an apparent magnitude in the B filter (effective wavelength 450\,nm; width 105\,nm) $m_B=18.67(3)$\,mag. However, another Gaia source (Gaia DR2 5946454411127231488) lies 0.8$\arcsec$ from Gaia4032, so we do not discuss this value any further. Given the OM PSF ($\sim1-2\arcsec$), the OM data are probably affected by flux contamination from the nearby source. In the left panel of Fig.~\ref{fig:lcfit}, we show a comparison between the EPIC pn+MOS light curve in the 0.3--10\,keV band of \src\ and the OM light curve obtained in fast mode of the optical counterpart. The two light curves share a similar modulating trend, especially in the second half of the observation. 

\subsection{Radio data analysis}

We detected a faint, polarised radio source at the location of \src\ in the time- and frequency-averaged ASKAP image, with the peak flux density $F_{\rm{rad}}=0.23\pm 0.04 \rm{\,mJy\,beam^{-1}}$. This is slightly lower than the continuum ASKAP detections reported by \cite{Rose2026}, though comparable to flux densities reported in that work for longer observations. For the distance range reported in \cite{Rose2026}, this corresponds to the specific luminosities of $L_{\nu,\rm{rad}}=10^{16}$--$10^{19}\rm{\,erg\,s^{-1}\,Hz^{-1}}$. We conducted Lomb-Scargle analysis using the time series data and identified periodic variability in the ASKAP data (see the inset in the middle panel of Figure~\ref{fig:lcfit}), despite the lack of any significant, bright pulses during the observation. This results in a maximum power peak at $P=4775(470)$\,s, which is consistent with the X-ray period and the radio period in \cite{Rose2026}.


\section{Discussion and conclusions}\label{sec:Discussion}

We have reported our radio, optical, and X-ray campaign of \src, an LPT recently discovered in archival ASKAP data and recognized as the first accreting CV showing bright polarized radio pulses \citep{Rose2026}. We reported on the significant detection of the 4868-s X-ray period in the \xmm\ and EP data, confirming \src\ as an accreting source. \cite{Rose2026} identified \src, the first LPT for which the radio periodicity is detected also in X-ray and optical data, as a binary system composed of a WD and an M dwarf. The presence of such a modulation in the X-ray data points to a magnetised WD. 

If interpreted as the spin period, such a long period suggests that \src\ is either a polar or a slowly rotating IP. Unfortunately, the length of our \xmm\ observation does not allow us to constrain the nature of the long-term trend, which would allow us to correctly classify \src. Assuming that $P_\mathrm{long}\simeq52$\,ks is a lower limit on the beat period and $P\simeq4868$\,s is the spin period, we can derive an upper limit on the orbital period of the system $P_\mathrm{UL}<5370$\,s, making \src\ an asynchronous polar. If \src\ is instead an IP, the 4868-s period would reflect the spin motion of the WD, while the orbital period would be much longer, possibly $P_\mathrm{long}\simeq52$\,ks. These long periodicities fall on the long end of the spin and orbital period ranges of IPs \citep[see, e.g.,][]{CotiZelati2016,Halpern2018}.

The \src\ flux during FXT4 and FXT5 (performed $\sim6$\,months after FXT3) is consistent with that observed during our 2025 campaign (Figure~\ref{fig:FXTevolution}), suggesting no large changes in the X-ray emission level. A strictly synchronous polar is expected to show a more drastic on-off behaviour \citep[see e.g. the reviews by][]{Cropper1990,Schwope2025}. A few asynchronous polars and IPs, on the other hand, have shown steadier fluxes \citep[see e.g.][]{Rea2017,Dutta2022}. 

The energy-dependent pulsed fraction, together with the hardness ratio showing a harder emission at the minimum of the pulse profile, is consistent with phase-dependent local absorption caused by an accretion curtain, as photoelectric absorption suppresses soft X-rays more efficiently than hard X-rays. A smooth sinusoidal modulation of the hardness ratio at the rotational period is not a characteristic of polars but rather of IPs and asynchronous polars \citep{Mukai2017}. Our spectral analysis also supports an IP or asynchronous polar interpretation. The spectral decomposition is consistent with the standard magnetic CV framework, with the hot, optically thin component associated with shock-heated plasma in the post-shock accretion column above the WD magnetic pole. The soft blackbody component, instead, can be interpreted as thermalised emission from the WD surface beneath the accretion column. The inferred emitting radius is smaller than the WD radius and can be associated with accretion-heated polar regions \citep[][]{Bernardini2012,deMartino2020}. We must note, however, that with the current uncertainty on the distance of \src\ any estimate of the radius of the blackbody component is to be treated with caution. The temperatures we derive for the spectral components are consistent with those observed in other IPs, although the blackbody temperature is particularly high \cite[see e.g.][]{Page2022}. The absorption feature at 0.77\,keV is also consistent at $3\sigma$ with absorption from O VII (0.74 keV), detected in other IPs \citep[see e.g.][and references therein]{Bernardini2012}.
 
In the left panel of Fig.~\ref{fig:LradioLXplane}, we show the position of \src\ in the $L_\mathrm{radio}-\LX$ phase space. \src\ is in the region of (magnetic) CVs. In the right panel, we show the position of \src\ in the $P_\mathrm{orb}-P_\mathrm{spin}$ plane. Given the uncertainties on the nature of $P_\mathrm{long}$, we show both possibilities: a) $P_\mathrm{long}$ is a (lower limit) on the orbital period; b) $P_\mathrm{long}$ is a beat period between the orbital and the spin period. \src\ is consistent either with (asynchronous) polars or IPs with long orbital periods. In summary, the available data support the classification of \src\ as a magnetic CV. Both the timing and spectral analysis point towards an IP or an asynchronous polar, but further observations are needed for a more definitive classification. In particular, longer uninterrupted X-ray observations, accompanied by high-cadence optical observations, will allow us to obtain a better estimate of the short-term period and to ascertain whether the long-term periodicity is caused by a beat.

\begin{acknowledgements}
    This work is based on data obtained with: \emph{Einstein Probe}, a space mission supported by the Strategic Priority Program on Space Science of the Chinese Academy of Sciences, in collaboration with ESA, MPE, and CNES (grant XDA15310000), the Strategic Priority Research Program of the Chinese Academy of Sciences (grant XDB0550200), and the National Key R\&D Program of China (grant 2022YFF0711500); \emph{XMM–Newton}, a European Space Agency (ESA) science mission with instruments and contributions directly funded by ESA Member States and National Aeronautics and Space Administration (NASA); Inyarrimanha Ilgari Bundara, the CSIRO Murchison Radio-astronomy Observatory. We acknowledge the Wajarri Yamaji People as the Traditional Owners and native title holders of the Observatory site. CSIRO’s ASKAP radio telescope is part of the Australia Telescope National Facility (\url{https://ror.org/05qajvd42}). Operation of ASKAP is funded by the Australian Government with support from the National Collaborative Research Infrastructure Strategy. ASKAP uses the resources of the Pawsey Supercomputing Research Centre. Establishment of ASKAP, Inyarrimanha Ilgari Bundara, the CSIRO Murchison Radio-astronomy Observatory and the Pawsey Supercomputing Research Centre are initiatives of the Australian Government, with support from the Government of Western Australia and the Science and Industry Endowment Fund. FCZ, MI, and NR are supported by the ERC Proof of Concept ``DeepSpacePulse'' (No. 101189496) and acknowledge funding from the Catalan grant SGR2021-01269 (PI: Graber/Rea), the Spanish grant ID2023-153099NA-I00 (PI: Coti Zelati), and by the program Unidad de Excelencia Maria de Maeztu CEX2020-001058-M, financed by MCIN/AEI/10.13039/501100011033, and by the MaX-CSIC Excellence Award MaX4-SOMMA-ICE. MI thanks Davide De Grandis for the discussion. FCZ is supported by a Ram{\'o}n y Cajal fellowship (grant agreement RYC2021-030888-I). ZW acknowledges support from the Australian Government through the Australian Research Council Discovery Project DP250102020 (PI: Hurley-Walker/Rea). DdM acknowledges support from INAF AF2022 FANS and AF2024 PULSE-X projects. DLK was supported by NSF grant AST-1816492. We thank the referee for the constructive comments.
\end{acknowledgements}

\bibliographystyle{aa} 
\bibliography{Bibliography/silmarillion} 

\begin{appendix} 
  \section{Data reduction}\label{appendix:datareduction}

  The source region for both \xmm\ and EP data was centered at the radio source coordinates ($\mathrm{R.A.}=17^\mathrm{h}45^\mathrm{m}08\fs929$, $\mathrm{Dec}=-50^\circ51'49\farcs856$, J2000) as derived by MeerKAT observations \citep{Rose2026}. Photon arrival times were corrected to the Solar System barycenter using the JPL DE405 ephemeris and the MeerKAT position.

    \begin{table}[htbp!]
    \centering
    \caption{Log of observations.
    }
    \resizebox{\columnwidth}{!}{
    \begin{tabular}{cccc}
    \hline
    \hline
     Instrument & ObsID & Start Date & Net exposure time\tablefootmark{a} 
     \\ 
      &  &  & (ks)\\ 
    \hline
    EP/FXT\tablefootmark{b} & 06800000900 & 2025 Sep 13 & 9.2 
    \\
    EP/FXT\tablefootmark{b} & 08500000410 & 2025 Sep 19 & 10.5
    \\
    EP/FXT\tablefootmark{b} & 08500000411 & 2025 Sep 23 & 10.0
    \\
    \xmm/EPIC & 0973390301 & 2025 Oct 04 & 34.1 / 44.7 / 44.7 
    \\
    \xmm/OM & -- & -- & 44.5
    \\
    ASKAP & 77513 & 2025 Oct 05 & 28.8
    \\
    EP/FXT\tablefootmark{b} & 11900670464 & 2026 Apr 03 & 9.0 
    \\
    EP/FXT\tablefootmark{b} & 11900711808 & 2026 May 02 & 9.8 
    \\
    \hline
    \hline
    \end{tabular}
    }
    \tablefoot{\xmm, \ep, and ASKAP observations analysed in this work.
    \tablefoottext{a}{In the case of \xmm/EPIC, the times are reported in the order EPIC pn/EPIC MOS1/EPIC MOS2. For both \xmm\ and \fxt, the times are those after cleaning from background flaring.
    }
    \tablefoottext{b}{In the text, we refer to these observations as observation FXT1, FXT2, FXT3, FXT4, and FXT5, respectively.}
    }
    \label{tab:XrayObs}
\end{table}

  \subsection{XMM-Newton}

We asked for an \xmm\ \citep{Jansen2001,Schartel2022} observation via Director Discretionary Time (DDT: ObsID 0973390301; PI: Rea). We considered data coming from the EPIC-pn camera \citep{Struder2001}, the EPIC-MOS cameras \citep{Turner2001}, and the Optical Monitor \citep[OM; ][]{Mason2001}. Reflection Grating Spectrometer \citep[RGS; ][]{denHerder2001} data were not considered for our analysis, since the low number of detected photons ($\simeq500$) in the background-subtracted, first-order spectra did not improve our spectral analysis. We performed data reduction and source events extraction in the 0.3--10\,keV band for timing and spectral analysis using \textsc{SAS} \citep{Gabriel2004} v22.1.0 with the latest \xmm\ calibrations and standard data reduction procedures. For the data reduction and event extraction of the EPIC-pn (EPIC-MOS) data, we used the \texttt{epproc} (\texttt{emproc}) task and selected only events with $\texttt{PATTERN}\leq4$ ($\texttt{PATTERN}\leq12$). We identified intervals of high-background particle flares by extracting the light curves of the entire field of view at energies $E>10$\,keV, filtering out intervals during which the pn (MOS) background count rate was $>0.4$ (0.35) cts/s. For source event extraction, we considered a circular region with a radius of 30\arcsec. A nearby, source-free circular region with radius of 50\arcsec\ on the same CCD was used for background extraction. We used the task \texttt{epiclccorr} to produce background- and vignetting-corrected light curves, while we created the response matrices and ancillary files using the tasks \texttt{rmfgen} and \texttt{arfgen}, respectively. We rebinned the energy spectra to have at least 25 counts per energy bin to apply $\chi^2$ statistics. 

Data from the XMM Optical Monitor (OM) were acquired both in Fast and Image modes with the B filter (effective wavelength 450\,nm, width 105\,nm). We used both the \texttt{omfchain} and the \texttt{omichain} tasks for the data reduction in the optical band.

\subsection{Einstein Probe}
The Follow-up X-ray Telescope \citep[FXT;][]{Chen2021,Chen2025,Zhang2025FXT} on board EP observed the source location of \src\ for three times in 2025 (PI: Rea, ObsIDs 06800000900, 08500000410, 08500000411) and twice in 2026 (PI: Wang, ObsIDs 11900670464, 11900711808), following the detection of its counterpart in the first FXT observation \citep{Rose2026} . In this letter, we refer to EP ObsIDs 06800000900, 08500000410, 08500000411, 11900670464, and 11900711808 as observations FXT1, FXT2, FXT3, FXT4, and FXT5, respectively. The data were reduced with the \texttt{fxtchain} tool available within the FXT Data Analysis Software Package (FXTDAS; \citealt{Zhao2025}). For timing analysis, we use the data collected in the full 0.3--10\,keV band of EP/FXT. However, for spectral analysis, we restricted our analysis to the 0.5--10\,keV band to avoid calibration issues. EP/FXT consists of two identical modules, FXT-A and FXT-B. During the five observations, both modules were set in the Full Frame mode (FF), with a time resolution of 50\,ms. A circular source region with a radius of $60\arcsec$ and an annular background region with inner and outer radii of $120\arcsec$ and $240\arcsec$ were adopted.

\subsection{ASKAP}

We requested a target of opportunity observation with ASKAP (SBID 77513) to be observed quasi-simultaneously with the \xmm\ DDT. We observed \src\ for $8$\,hr at a central frequency of $1365$\,MHz, with $144$\,MHz bandwidth\footnote{The bottom half of the $288$\,MHz-wide ASKAP-mid band is heavily affected by radio-frequency interference}, beginning at 02:58 UTC on October 5 2025. The data were calibrated with the standard ASKAP pipeline \citep[see][]{Hotan2021}. We extracted model-subtracted dynamic spectra and formed folded radio lightcurves using the \textsc{DStools} package \citep{dstools}, following the procedure described in \citep{Rose2026}.

\section{XMM-Newton timing analysis bootstrapping and plots}\label{appendix:boot}

\begin{figure}
    \centering
    \includegraphics[width=0.95\columnwidth]{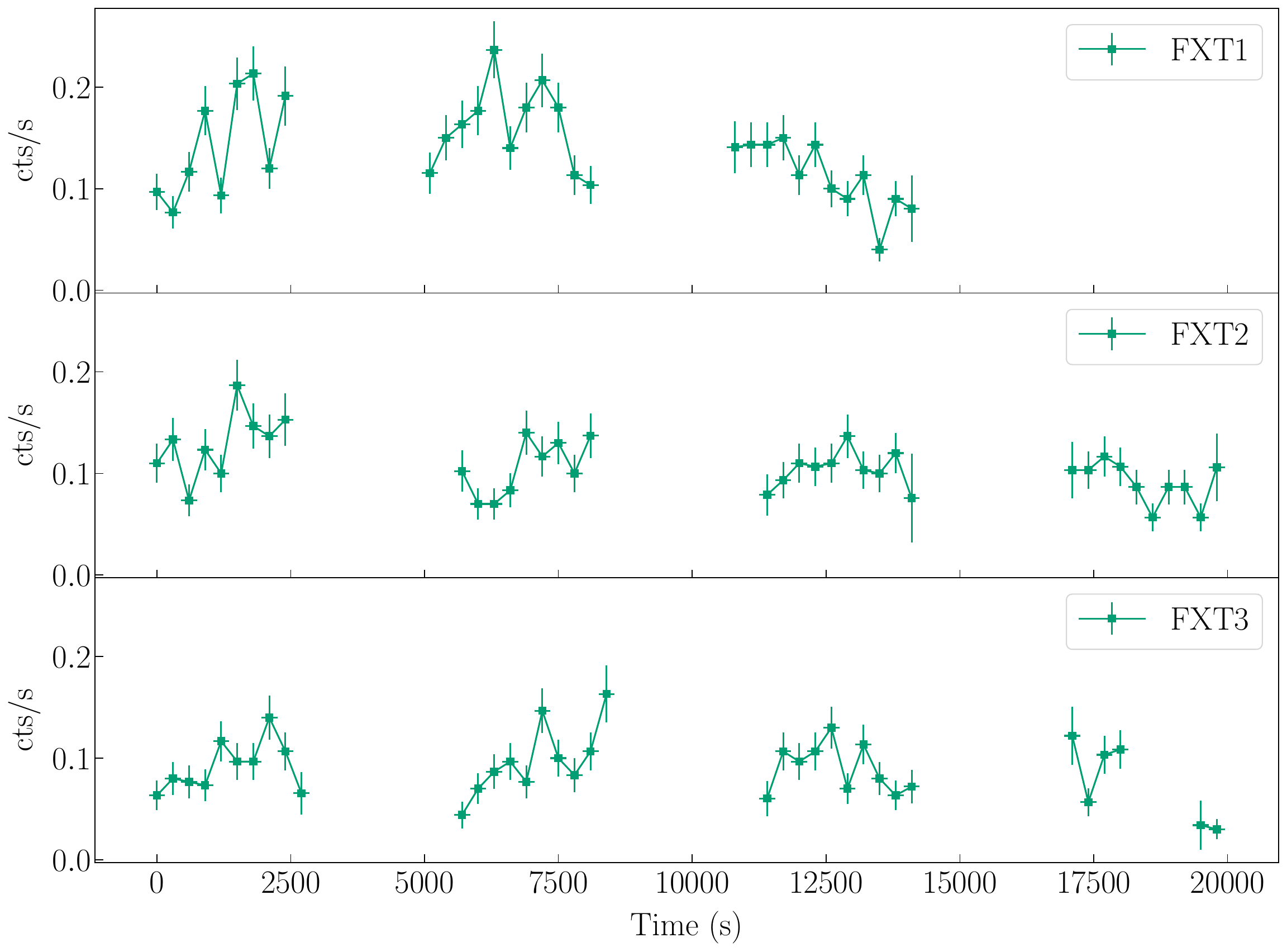}
    \caption{EP/FXT light curve in the 0.3--10\,keV band during observation FXT1 (top panel), FXT2 (middle panel), and FXT3 (bottom panel).}
    \label{fig:pds}
\end{figure}

Given the low frequency of the peak detected in the PDS, we adopted the algorithm proposed by \cite{Israel1996} to estimate the significance of a PDS peak in the case of red noise. Although the peak itself is not significant and has a power corresponding to $\simeq87\%$ of the 3$\sigma$ detection threshold, we note that the peak frequency is consistent with the $1/P_\mathrm{radio}$ frequency derived by \cite{Rose2026}. 

\begin{figure}
    \centering
    \includegraphics[width=0.95\columnwidth]{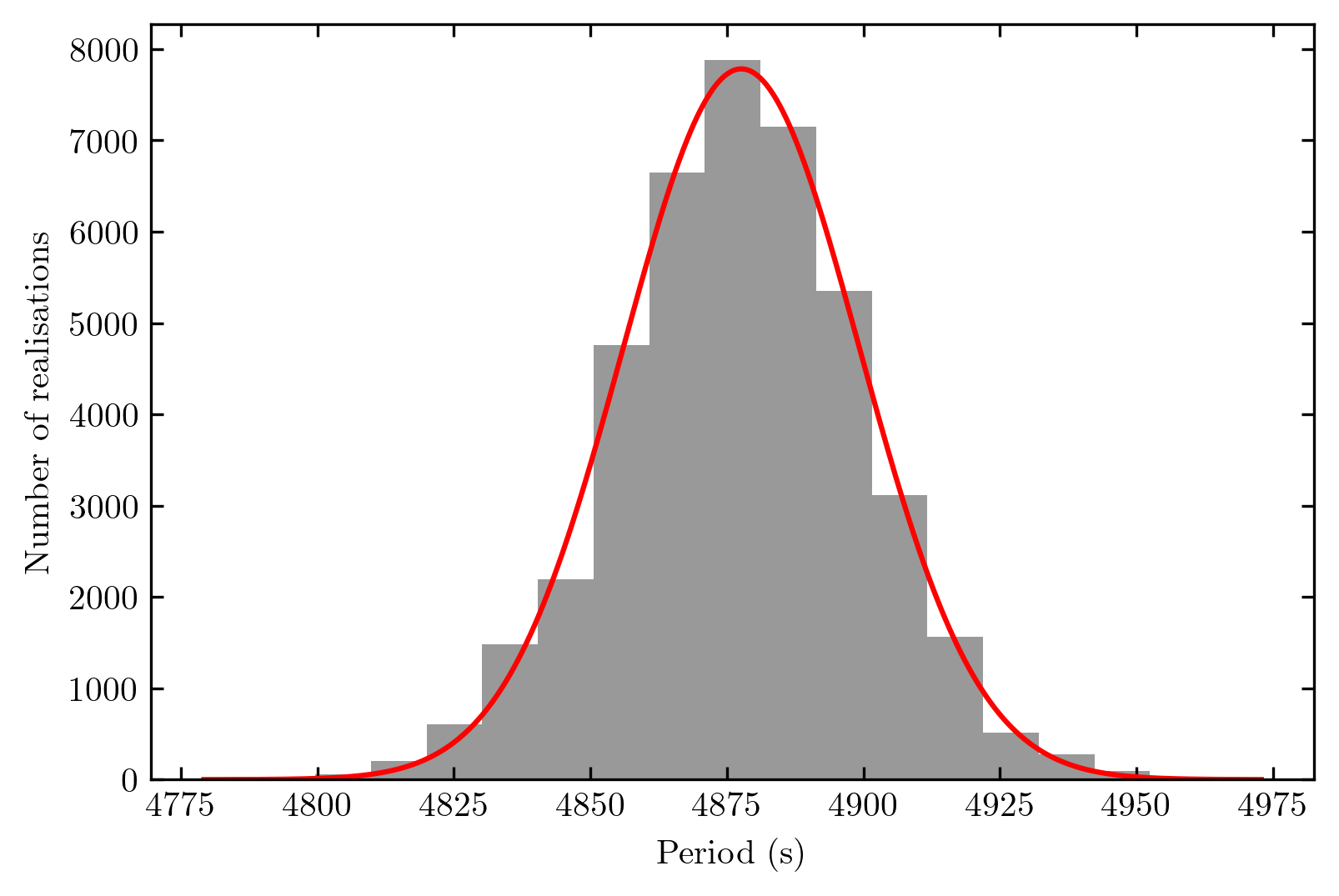}
    \caption{Distribution of the centroid period found by bootstrapping. In grey, the \xmm\ EPIC-pn data, with the Gaussian fit (in red) peaking at 4878(22)\,s.}
    \label{fig:bootstrap}
\end{figure}

We estimated the period observed in the X-ray data and its uncertainty through bootstrapping, as applied in \cite{Paice2024MNRAS.531L..82P} and \cite{Veresvarska2025a}.
We sampled the EPIC-pn light curve $5\times10^{4}$ times with replacement and computed a Lomb-Scargle periodogram \citep{Lomb1976,Scargle1982} of the new light curve with frequency grid oversampled by a factor of 4 \citep[as done in][]{Veresvarska2025a} and constrained to the immediate vicinity of the signal. Lomb-Scargle periodogram is used to account for the uneven sampling created by resampling. The peak of the detected signal is recorded and the resulting distribution is shown in Fig.~\ref{fig:bootstrap}. The centroid period is recovered through a Gaussian fit to the distribution resulting in a period of $P=4878(22)$\,s. We note that the peak frequency during the bootstrapping process was recovered in $\sim84\%$ cases of the $5\times10^{4}$ realisations, which is consistent with the previously determined significance of the signal. A similar separate search on the \fxt\ data from observations FXT1, FXT2, and FXT3 alone results in a period estimate of $P=4840(120)$\,s. The larger errors are probably caused by the presence of many gaps in the \fxt\ light curve (Figure~\ref{fig:pds}), resulting in a windowing effect that makes the detection of the signal harder. We then refined the \xmm\ bootstrapping value through epoch folding and phase-fitting with the \fxt\ data, obtaining a period $P=4868(22)$\,s. The $1\sigma$ uncertainty derived by phase-fitting is 0.2\,s. The long baseline (approximately 20 days) between the first \fxt\ observation and the \xmm\ observation was crucial to derive the phase-fitting solution. However, the associated error is most likely underestimated, dominated by the baseline itself and not a reflection of the true uncertainty associated with this measure. Therefore, we chose to retain the more conservative error derived from bootstrapping. 

\begin{figure}
    \centering
    \includegraphics[width=0.95\columnwidth]{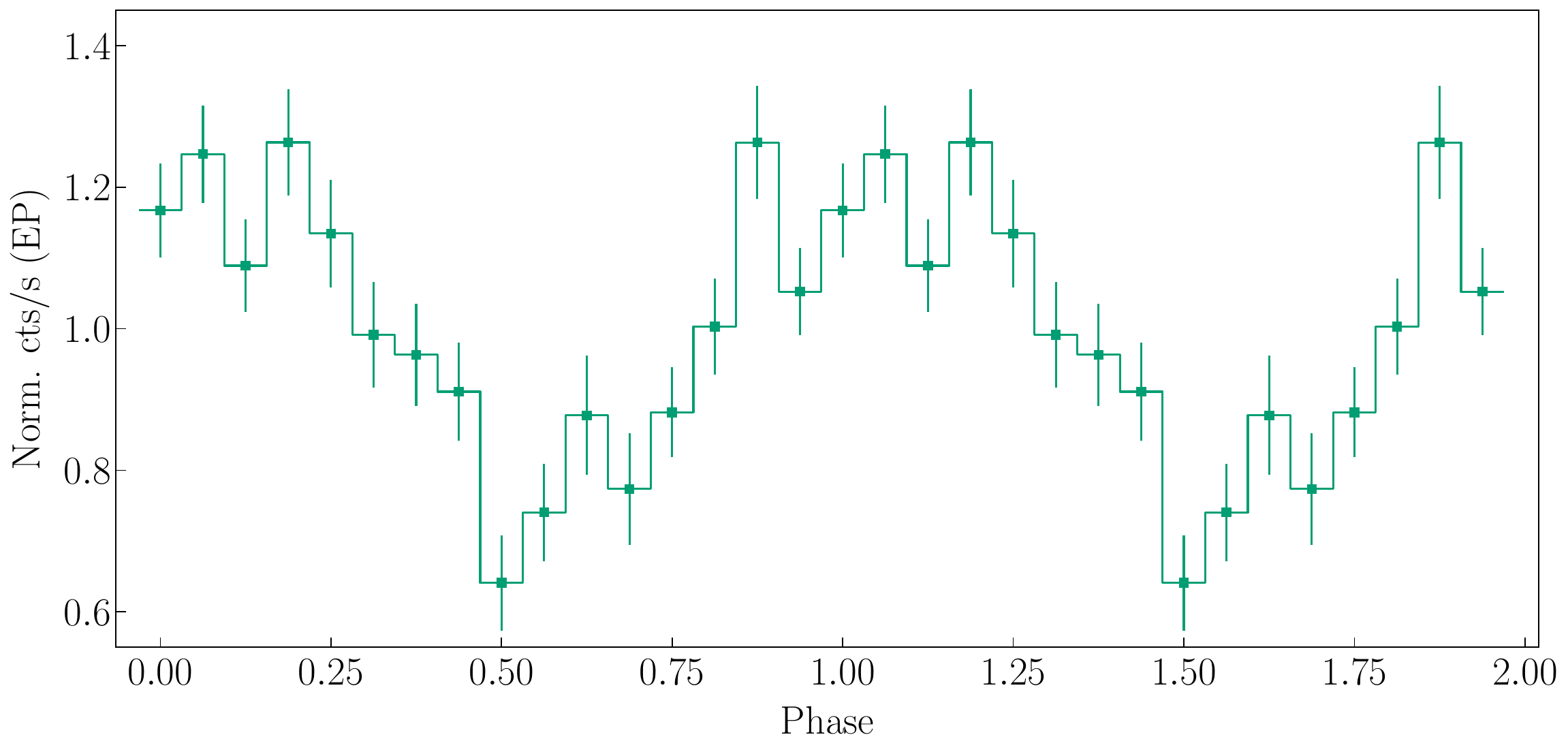}
    \caption{EP folded light curve ($P=4868$\,s, reference epoch MJD 60952) of observations FXT1/2/3.}
    \label{fig:epprofile}
\end{figure}

\section{Additional XMM-Newton/EPIC and EP spectral analysis}

Apart from the model shown in this Letter, we also considered other combinations of components (e.g., \texttt{bbodyrad+powerlaw}, \texttt{apec+apec}, \texttt{mekal+mekal}, or \texttt{cempow+bbodyrad}), but all other models resulted in worse fits. Nonetheless, it is interesting to note that, whenever two thermal components were included in the model, the derived temperatures were always consistent with the one reported in Table~\ref{tab:spec}.

 \begin{table}
\caption{\xmm\ spectral analysis of \src.
}
\centering
\resizebox{\columnwidth}{!}{
\begin{tabular}{ l l l l }
\hline
\hline
\multicolumn{4}{c}{\bf \xmm\ spectral analysis}\\
\hline
\multicolumn{4}{c}{{\bf model:} \texttt{constant}$\times$\texttt{tbabs}$\times$\texttt{tbpcf}$\times$\texttt{gabs}$\times$(\texttt{apec}+\texttt{bbodyrad})} \\
\hline
Component & \multicolumn{2}{c}{\bf Parameters} & Value\\
\hline

\multirow{3}{*}{\texttt{constant}}
& $c_\mathrm{PN}$ &
& (1.0)\\

& $c_\mathrm{MOS1}$ &
& $0.97\pm0.03$\\

& $c_\mathrm{MOS2}$ &
& $0.93\pm0.03$\\

\hline

\texttt{tbabs}\tablefootmark{a}
& $N_\mathrm{H}$ & 10$^{22}$ cm$^{-2}$
& $(0.15)$\\

\hline
\multirow{3}{*}{\texttt{tbpcf}}
& $N_\mathrm{H}$ & 10$^{22}$ cm$^{-2}$
& $0.8\apm{0.2}{0.5}$\\

& Covering fraction & 
& $0.8\apm{0.3}{0.1}$\\

& $z$ &
& (0) \\

\hline
\multirow{3}{*}{\texttt{gabs}}
& $E_\mathrm{line}$ & keV
& $0.77\pm0.01$\\

& $\sigma$ & keV
& $0.002\apm{0.0006}{0.03}$\\

& Normalization &
& $>4$\\

\hline
\multirow{4}{*}{\texttt{apec}}
& $kT_\mathrm{apec}$ & keV
& $15\apm{3}{4}$\\

& Abund & Solar
& (1.0)\\

& $z$ &
& (0) \\

& Norm & $10^{-3}$
& $1.02\pm0.03$\\

\hline
\multirow{2}{*}{\texttt{bbodyrad}\tablefootmark{b}}
& $kT_\mathrm{bb}$ & keV
& $0.14\apm{0.01}{0.03}$\\

& $R_\mathrm{bb}$ & km
& $1.0\apm{0.7}{0.6}$\\

\hline
\multirow{2}{*}{\texttt{cflux}\tablefootmark{b,c}}
& $F_\mathrm{0.5-10\,keV}$
& 10$^{-12}\ergcms$ 
& $2.5\pm0.3$ \\
& $L_\mathrm{0.5-10\,keV}$ & 10$^{32}\ergs$ 
&
$1.0\pm0.1$
\\

\hline
& $\chi^2$ & (dof) & 366.57/335
\\

\hline
\hline
\end{tabular}
}
\tablefoot{
    The errors show the 90\% confidence level. We froze the parameters in parentheses.
    \tablefoottext{a}{Fixed at the average value of the Galactic absorption along the line of sight between 0.57 and 6.5\,kpc, as computed by the 3D NH-tool \citep{NHTool2024}.}
     \tablefoottext{b}{$d\simeq0.57$\,kpc, as measured by \textit{Gaia} parallax.}
     \tablefoottext{c}{Unabsorbed flux and luminosity in the 0.5--10\,keV band.}
     }
\label{tab:spec}
\end{table}

\section{Additional XMM-Newton/OM analysis}

We also computed the cross-correlation function to verify the presence and estimate lags between the two bands. However, with the available dataset no significant features were visible. We also searched for periodic signals in the OM data in PDSs computed with bin times of 10\,s (the smallest available for the OM data) and 300\,s, but no signal was detected, in particular at the period detected in the X-ray and radio bands.

\section{\src\ comparison with other sources}

\begin{figure*}
    \centering
    \includegraphics[width=0.95\columnwidth]{Plots/radio_Xray_ASKAPJ1745_new.pdf}
    \includegraphics[width=0.95\columnwidth]{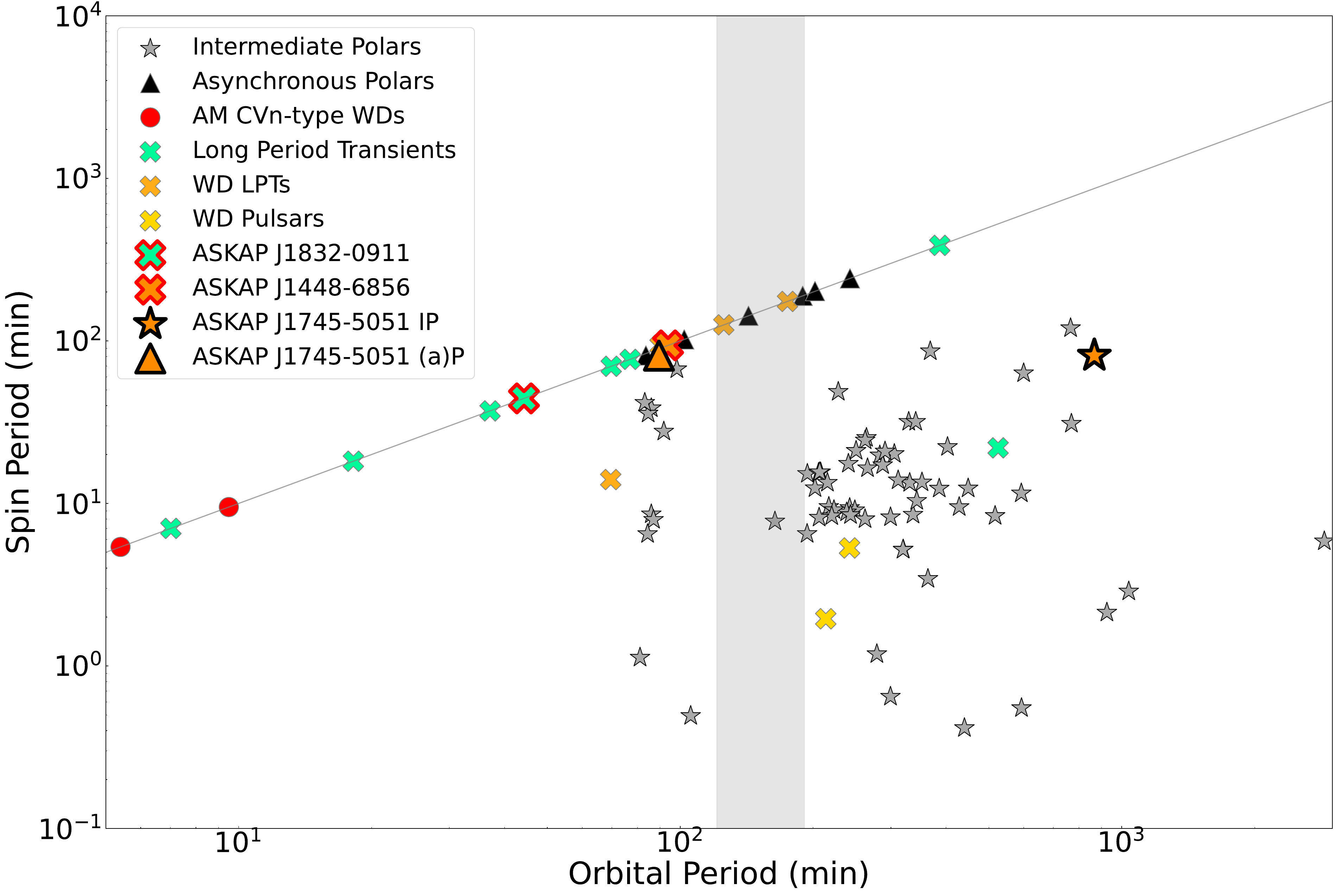}
    \caption{\textit{Left}: \src\ position in the $L_\mathrm{radio}-\LX$ plane, shown by the orange cross with black edge. The other classes of sources shown in the plot are described in the plot itself. \textit{Right}: \src\ position in the $P_\mathrm{orb}-P_\mathrm{spin}$ plane, compared to IPs, (asynchronous) polars, LPTs, WD pulsars and other WD LPTs. We show \src's position both in the IP scenario (orange star with black edges) and the (asynchronous) polar scenario (orange triangle with black edges). The plots are adapted from \cite{Wang2025}, \cite{Rea2026}, and \cite{Mukai2017}. In both plots, we also highlight the other two LPTs detected in X-rays, that is ASKAP\,J1448-6856 and ASKAP\,J1832-0911. In the right plot, the grey solid line shows the values for which $P_\mathrm{spin}=P_\mathrm{orb}$, while the grey shaded area shows the CV ``period gap''.}
    \label{fig:LradioLXplane}
\end{figure*}

\end{appendix}

\end{document}